\documentclass[aps,prx,amssymbs,showpacs]{revtex4}
\usepackage{epsfig,times,xcolor}
\usepackage{amsmath, mathrsfs}

\begin{document}
\title{Elasto-buoyant heavy spheres: a unique way to study non-linear elasticity}
\author{Aditi Chakrabarti}
\affiliation{Department of Chemical and Biomolecular Engineering. Lehigh University, Bethlehem, Pennsylvania 18015, United States}

\author{Manoj K. Chaudhury}
\affiliation{Department of Chemical and Biomolecular Engineering. Lehigh University, Bethlehem, Pennsylvania 18015, United States}

\author{Serge Mora}
\email[]{serge.mora@umontpellier.fr}
\affiliation{Laboratoire de M\'ecanique et de G\'enie Civil. UMR 5508, Universit\'e Montpellier and CNRS. 163 Rue Auguste Broussonnet. F-34090 Montpellier, France.}

\author{Yves Pomeau}
\affiliation{University of Arizona, Department of Mathematics, Tucson, USA.}

\date{\today}
\begin{abstract}
   Extra-large deformations in ultra-soft elastic materials are ubiquitous, yet systematic studies and methods to understand the mechanics of such huge strains are lacking. Here we investigate this complex problem systematically with a simple experiment: by introducing a heavy bead of radius $a$ in an incompressible ultra-soft elastic medium. We find a scaling law for the penetration depth ($\delta$) of the bead inside the softest gels as $\delta \sim a^{3/2}$ which is vindicated by an original asymptotic analytic model developed in this article. This model demonstrates that the observed relationship is precisely at the demarcating boundary of what would be required for the field variables to either diverge or converge. This correspondence between a unique mathematical prediction and the experimental observation ushers in new insights into the behavior of the deformations of strongly non-linear materials.

\end{abstract}

\maketitle

\begin{figure}[!h]
\begin{center}
\includegraphics[width=0.55\textwidth]{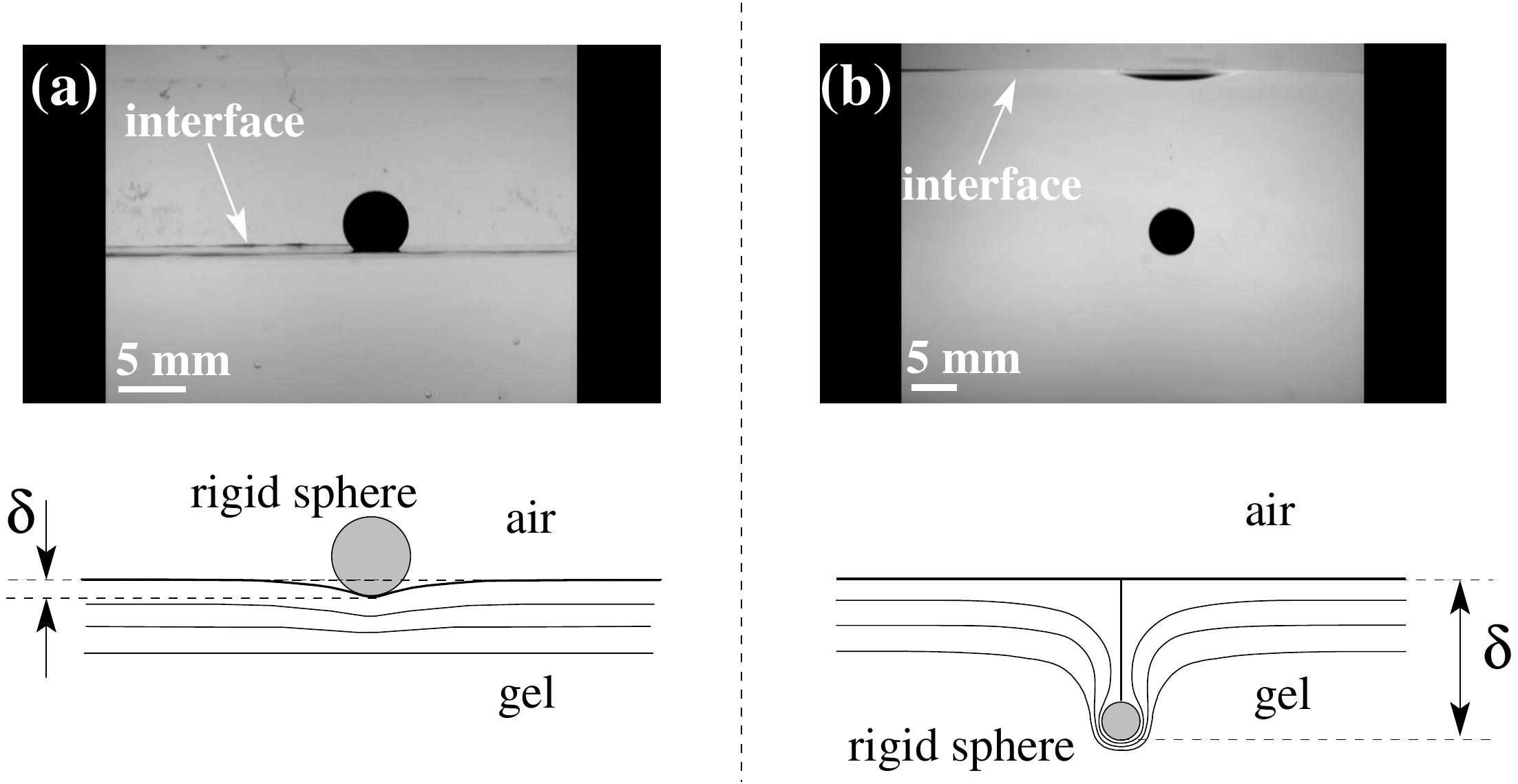}
\end{center}
\caption{Snapshots and schematics of the experiment. ({\bf Top}) Side views of two transparent cells filled with a polyacrylamide 
  gel having a shear modulus of 1160 Pa (a) and  13 Pa (b). Two identical steel beads 
  (5 mm diameter) have been deposited in the free air-gel interface. The vertical 
  downshifts are respectively $\delta=0.03~\delta_0$  and $\delta=320~\delta_0$. ({\bf Bottom})  Schematic of the experiments. } 
\label{fig : schematic}
\end{figure}

Singularities are pervasive in various problems of linear continuum mechanics. In wetting, 
stress diverges at a moving contact line \cite{Degennes1985,Pomeau2002}; it diverges at the tip of a crack 
or even at a sharp point indenting a plane \cite{Johnson1985}. Understanding how such singularities can be tempered has often given rise to new physics invariably prompting us, on many occasions, to investigate a material phenomenon at a molecular dimension and then herald a way to bridge the near field 
with the far field behavior in a rather non-trivial manner.
Nature, however, performs the difficult task herself and leaves her signature in a way that is independent of the constitutive property of a material, yet it belongs to a class of universality. What we report here is such a universality that is discovered in the large deformation behavior of ultra-soft gels. Soft solids undergoing huge deformations exhibit various fascinating and puzzling mechanical behaviors \cite{Hayward2010,Mora_prl2010,Mora_softmatter2011,Style2013,Saintyves2013,Mora_prl2013,Jagota2012,Chaudhury2015,Tallinen2014,Zhao2010,Chakrabarti2013a,Crosby2015}. Our experimental protocol to study extra-large elastic deformations is remarkably simple, in that a heavy bead of stainless steel is gently deposited on the horizontal flat surface of a gel. The compliant gel is deformed by the load exerted by the heavy bead. It reaches a stable (elasto-buoyant) equilibrium position when the elastic force exerted by the surrounding gel balances its weight \cite{Chakrabarti2013a}, within few tenths of a second. 
This experiment can be viewed as an elastic analog of the falling ball viscometry, in which the bead reaches a terminal sedimentation velocity resulting from the balance of the bead's weight and the viscous drag force \cite{Stokes1845}. In the limit of Hookean elasticity, an analogy with the Stokes equation (by replacing shear viscosity with shear modulus and velocity with depth of submersion, $\delta$)\cite{Rayleigh1922} suggests that $\delta \sim a^2$, $a$ being the sphere radius. While for the higher modulus gels (Fig.\ref{fig : schematic}-a), such a relationship is more or less valid, for the softer ones, when the bead is totally engulfed by the gel and the deformations are very large (Fig.\ref{fig : schematic}-b), it is observed that the depth scales with the bead's radius with an exponent of 3/2. This is a non-trivial result that cannot be explained by the usual neo-Hookean model, i.e, by considering a quadratic elastic energy density with respect to finite strains. According to this model, the displacement would be infinite! When the same problem is analyzed using a new analysis presented in this paper,
it is found, remarkably, that the exponent of the radius of the bead is just at the juncture of 
what would be required for the field variables to avoid divergence. What is quite remarkable about this analysis is that no detailed non-linear behavior of elastic solids \cite{Rivlin1948,Ogden1972,Yeoh1993,Gent1996} needs to be speculated. The scaling laws are universal that being independent of the particular constitutive law of the elastic material. This unique correspondence between a mathematical prediction and the experimental results, thereby unfolding new physics of highly non-linear deformations, is the subject of this paper.

\section{Experimental evidence of elasto-buoyancy}
Cross-linked polyacrylamide gels are used in the experiments reported below. The gel solutions are prepared according to the recipe published previously \cite{Chakrabarti2013a,Chakrabarti2014} and cured in home-built glass containers
(70 mm x 50 mm x 40 mm). The inner walls of the containers are grafted with a thin layer ($\sim$5 nm) of
polydimethyl siloxane chains so that the gel solution contacted
the walls at 90$^o$ to ensure that the surface of the cured gel is flat.
All the experiments are performed
after two hours of gelation. For the estimation of the shear modulus of the gel, we use a linear elastic model in order to ensure consistency of quantification for all the gels. The shear modulus is determined from the resonant
mode of vibration of a gel slab confined between two parallel glass slides
\cite{Chakrabarti2014}. Steel spheres (density 7.8 g/cc, diameters 1 - 10 mm) are gently placed  one by one
on the gel surface and its side-view image is captured by a camera. Dissipative processes within the gel dampen any oscillations,
and the spheres sink until they 
become stagnant in the polyacrylamide gel, the whole process taking only fractions of a second. The depth of submersion, $\delta$, is measured
from the upper surface  of the gel till the base of sphere, that denoting the net downward
displacement due to the inclusion of the spherical particle by the surface
(Fig.\ref{fig : schematic}). The cells are large enough to avoid any finite boundary effects, including the side walls and the bottom. The measurement of depth for each sphere is made in the central region of the container. After each measurement, the sphere is gently removed from the gel using a magnet, held slightly away from the free surface. We wait for a few minutes between each measurement that ensures there is no memory of the position of the previous sphere inside the gel.

If the bead is too small and the gel is stiff \cite{Rimai2000,Style2013}, the surface bends slightly under the weight of the bead and $\delta \ll a$ (Fig.\ref{fig : schematic}-a).  By increasing the bead radius or decreasing the elastic modulus, the particle submerges itself to a
considerable depth inside the  gel. The surface of the gel wraps around the particle and
closes to create a line singularity  connecting the particle to the free surface of the gel
(Fig.\ref{fig : schematic}-b). Strings of tiny air bubbles appear in this thin channel, which
soon coalesce and escape through it while the channel further closes due to the auto-wetting
forces of the gel's surface. If the surface of the gel is premarked with ink spots, it is easy
to visualize that the surface of the gel becomes appreciably stretched while the sphere sinks
through the gel while being still connected to the free surface via a thin channel.
It is also possible to release ink inside the gel in the form of thin vertical lines with the
help of a fine needle, which bend toward the sphere in a dramatic way when the sphere is released
inside the gel demonstrating a substantial amount of tensile strain developed in the gel network due to the inclusion of the bead. These basic experiments were reported in a previous article \cite{Chakrabarti2013a}, but without a detailed analysis.
Here we report a detailed set of experiments, in which the shear modulus of the gel was varied from 13 Pa to about 3000 Pa.\\

\begin{figure}[!h]
\begin{center}
  \includegraphics[width=0.6\textwidth]{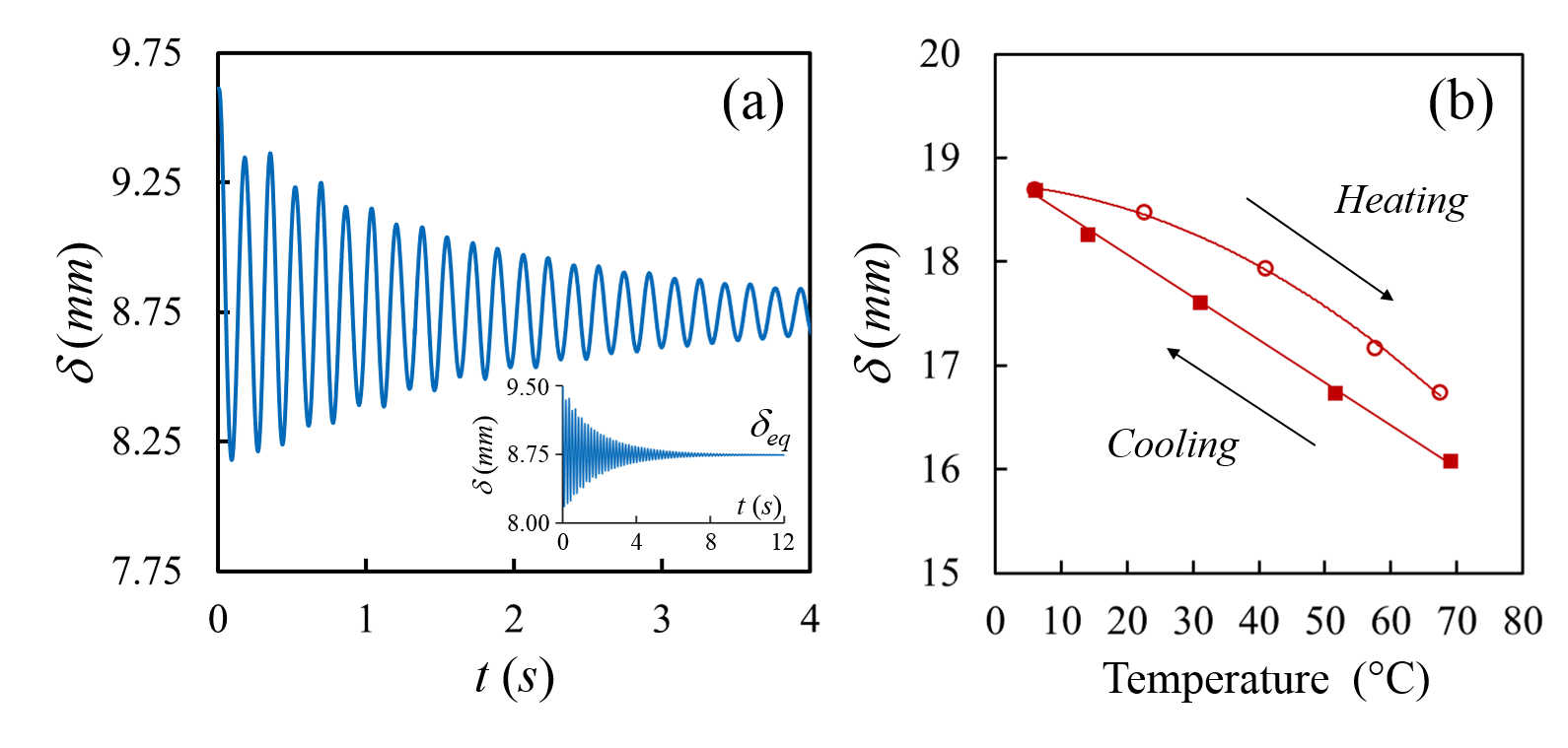}
\end{center}
\caption{Experimental evidence of reversibility of the gels. {\bf (a)} A 10 mm diameter steel sphere immersed in 140 Pa gel, when slightly disturbed from its elastobuoyant position via an electromagnet, undergoes under-damped oscillations about its equilibrium depth ${{\delta}_{eq}}$. {\bf (b)} Depth of
  submersion $\delta$ of a 5 mm diameter steel sphere in a soft gel ($\mu \sim$ 13 Pa) varying as a function
  of its temperature. The experiments in the cooling cycle were performed first
  following which the gel was heated systematically to obtain the data for the heating cycle.}
\label{fig : reversibility}
\end{figure}

Prior to subjecting these experimental results to a comparison with a theoretical analysis, we needed to verify how meaningful it is to consider only the effect of elasticity by ignoring the surface tension of the gel  in predicting the depth of submersion of the bead and how reversible is the deformation of the gel. First question is partly philosophical that rests upon the distinction between surface free energy $\gamma$ and surface tension. The latter differs from the former by surface stress $d\gamma/de$, $e$ being the surface strain. As the major constituent of these amorphous gels is water, we expect that the surface stress is negligible. In fact, several recent studies that measured surface tensions of various amorphous soft polymers strongly suggest that their surface tensions are practically same as their surface free energies  \cite{Ghatak2015}. Thus we need to figure out only if the surface free energies of the gels play any role. Part of the answer can be obtained by comparing them with the equivalent spring constant of the sample. The latter can be obtained by slightly raising the height of the bead by an electromagnet and releasing it so that the bead undergoes an under-damped oscillation  and reaches the neutral position. For three gels, in which the beads were completely submerged, the equivalent spring constants were estimated from the frequency of oscillation to be 0.2 N/m, 5.6 N/m and 13 N/m, which  increase systematically with their shear moduli (13 Pa, 140 Pa and 360 Pa). Fig.\ref{fig : reversibility}-a shows a typical profile of such an oscillation. Comparing these spring constants with the surface tension (0.07 N/m) of water, we conclude that the contribution of surface tension can be safely neglected for all the gels used in this study except, perhaps, for the lowest modulus gel for which the spring constant is three times that of the gel's surface tension. However, when  a bead is completely submerged in the gel, any variation of the height of the sphere does not alter the area and the excess energy of the free surface of the gel. We thus believe that the surface tension spring does not play a significant role in determining the depth of submersion of the sphere as long as it is completely engulfed by the gel. Further support to this viewpoint, i.e. the dominant role of the elasticity over surface tension was gathered from the experiment described below that also exemplified the reversibility of the softest gel ($\mu \sim$ 13 Pa) employed in our experiments. \\
 
By varying only the system temperature, the elastic modulus of polyacrylamide gels changes, which increases with temperature. A thin layer of Paraffin oil is poured over the gel to avoid its surface from drying.  The temperature of the gel is monitored by placing a thermometer inside an identical sample of gel in a similar sized container, placed inside the oven. After the gel is heated to 70 $^o$C, a 5 mm diameter steel sphere is released into it through the layer of paraffin oil. The depth of the sphere is measured at this temperature while it is in the oven. As the gel is gradually cooled, the sphere sinks deeper inside it. We wait for an hour between acquiring data for the depth of sphere at each temperature to allow the gel to equilibrate reasonably well. After the gel is cooled to about 5 $^o$C, it is heated again that decreases the depth of the embedded steel sphere (Heating Cycle, Fig.\ref{fig : reversibility}-b). The depth of submersion of the sphere plotted as a function of the temperature for both the cooling and the heating cycles (Fig.\ref{fig : reversibility}-b) shows that there is a little hysteresis in this system, in that the difference in the depths of the sphere for a given temperature is within 5\%. We conclude that the deformations of the gel generated by a bead are predominantly reversible and controlled by elastic forces.\\

\begin{figure}[!h]
\begin{center}
  \includegraphics[width=0.6\textwidth]{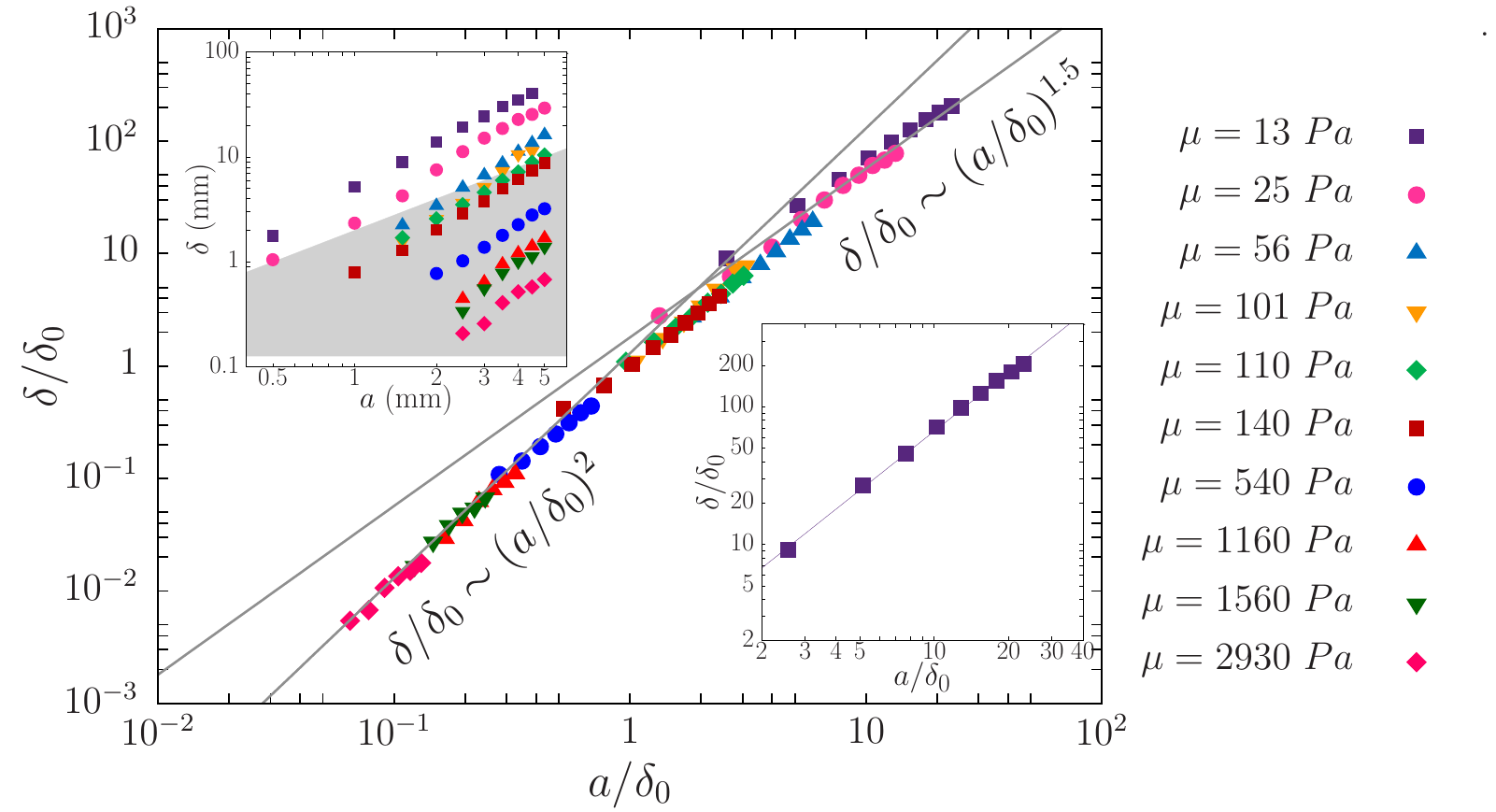}
\end{center}
\caption{Dimensionless depth of spheres ($\delta/\delta_0$) plotted  as a function of its dimensionless radius $a /\delta_0$, for various shear modulus of the gels from $\mu =$ 13 Pa to 2930 Pa. The grey lines indicate power law curves ($\delta/\delta_0~(a/\delta_0)^\alpha$) with $\alpha=2$ and $\alpha=1.5$ for the two asymptotic limits for the normalized data. (Upper Inset). Depths ($\delta$) versus radii ($a$) for all the spheres. The plot area is divided into two domains; the boundary indicating $\delta=2a$. The data points above the boundary indicates that the spheres were entirely  below the gel's surface. (Lower Inset). Best fit ($\delta/\delta_0=k(a/\delta_0)^\alpha$) for the softest gel (13 Pa) highlighted (see text).} 
\label{fig : data}
\end{figure}

The penetration depth, $\delta$, is plotted versus the radius $a$ of the steel spheres in log-scales in upper inset of Fig.\ref{fig : data}, for ten different gels ($\mu:$ 13 Pa - 2930 Pa) 
and different radii. The upper white region of this inset shows the data points for beads that were completely below the surface of the gel. When we non-dimensionalize the depths as well as the radii of each bead in each gel by the material lengthscale $\delta_0$ defined as $\delta_0 = \frac{\mu}{\Delta \rho g}$(Fig.\ref{fig : data}), where $\mu$ is the shear modulus of the gel, $\Delta \rho$ is the effective density of the buoyant spheres, and $g$ is the gravitational acceleration, we see that all the data cluster around a mean master curve with two distinctive asymptotic limits. For each sphere-gel system, $\Delta \rho$ was precisely estimated from the image on the basis of the Archimedes principle of buoyancy analogous to the scenario for floatation on a liquid \cite{Vella2015}. The normalized data can be divided more or less into two regimes: one with the non-dimensional radii $a/\delta_0 < 1$ and other one with $a/\delta_0 > 1$ with an intermediate transition regime. We fit the non-dimensional depths as a function of the non-dimensional radii for the regime $a/\delta_0 < 1$ with the power law function $(\delta/\delta_0)=k(a/\delta_0)^\alpha$ with adjustable parameters, $\alpha$ and $k$ and find $\alpha= 1.96 \pm 0.06$ and $k= 1.09 \pm 0.1$. The error bars are obtained from a $95\%$ confidence limit analysis. We conclude that for $a/\delta_0 < 1$, the depths as a function of the radii follows $\delta \sim a^2$ within the error limits. On careful examination of the experimental points corresponding to large deformations, i.e. $a > \delta_0$, we find that the data corresponding to the two softest gels ($\mu : $ 13 Pa and 25 Pa) are shifted from the rest of the data due to their multiplicative factor ($k$) being significantly larger than the rest of the data for the other gels. This indicates that the lengthscale $\delta_0$, which is defined with the elastic linear properties of the sample, is therefore not sufficient to describe the whole data accurately, and non linear effects have to be taken into account. We conclude that all the data in this regime ($a/\delta_0 > 1$) cannot be investigated together. It is more appropriate to investigate the data for each gel composition separately {\em i.e,} for a given non-linear material stress-strain relationship. By fitting the data for the two softest gels ($\mu : $ 13 Pa and 25 Pa) where all the beads are completely engulfed, with the power law function, we find that $\alpha_{13~ Pa}= 1.42 \pm 0.05$ (see lower inset of Fig.\ref{fig : data}), $k_{13~ Pa}= 2.52 \pm 0.3$ and $\alpha_{25~Pa}= 1.52 \pm 0.09$, $k_{25~ Pa}= 1.63 \pm 0.3$. The exponents for the fits for the gels ($a/\delta_0 \sim 1$) in the intermediate regime lie between $\alpha \sim 1.5$ and $\alpha \sim 2$. Thus, from the experimental observations, we infer that in the limit of $a$ significantly greater than $\delta_0$, the general trend is close to $\delta \sim a^{p}$, where $p$ is in the range of 1.4-1.5, within the error limits. 
Thus, we conclude that the power law observed ($\delta \sim a^2$) in the gels of higher shear moduli is closer to the regime  already studied before what one would expect with an analogy with the elastic Stokes equation in the Hookean limit. What is astounding is the observation of $\delta \sim a^{1.5}$ in the case of the gels where $\delta > 2a$ for each tested bead radius. In order to interpret these novel observations in the ultra-soft gels, we develop a new model to tackle such extra-large deformations in the following part of the paper.\\

\section{Asymptotic analytic model for large elastic deformations}
 
From  the outset, a motivating picture of the problem can be gleaned from the comparison of the potential energy of the bead in the gravitational field and the energy of the elastic deformation of the gel. In the limit of Hookean deformation, an analogy with the Stokes equation suggests that the elastic energy is of the form $\mu{\delta}^2a$, which is to be compared with the gravitational potential energy $\Delta\rho g a^3\delta$. This leads to the scaling: $\delta \sim a^2$. However, the experimentally observed scaling $\delta\sim a^{3/2}$ implies that the elastic energy in extremely large deformation must scale as $\delta^3$. The detailed analysis based on a model, presented in this paper for the first time, shows that the above scaling is non-trivial. Moreover, the scaling result that follows is independent of the constitutive laws of an elastic material.

\subsection{Gravity energy of engulfed spheres}
Consider deformations that preserve the volume, as is the case with the elastic
gels used in the experiments. Since the bead is supposed to be totally engulfed
with $\delta \gg a$, the surface of the deformed gel is fairly flat and
horizontal. A supplementary (virtual) downshift $\delta'$  produces an opposite
rise of the same volume of gel. Therefore the gravitational energy variation of the
system consisting of the sphere plus the gel is $\frac{4}{3}\pi a^3 \Delta \rho g\delta'$. We conclude
that the gravitational energy shift is
${\cal E}_{gr}\simeq \frac{4}{3}\pi a^3 \Delta \rho g \delta$ in the limit we are considering.

\subsection{Outline of calculating the elastic energy in the limit $\delta \gg a$}
\begin{figure}[!h]
\begin{center}
\includegraphics[width=0.5\textwidth]{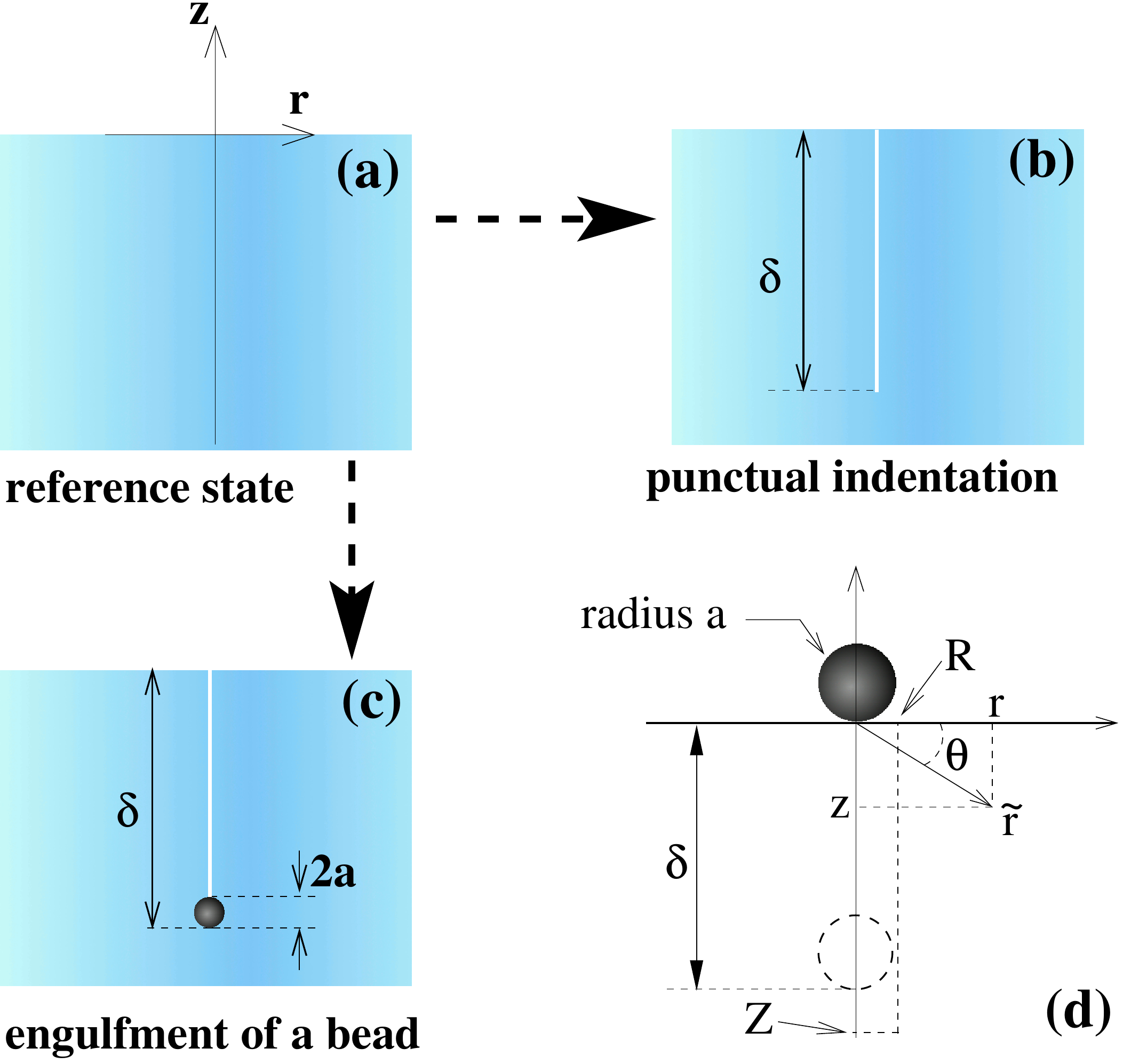}
\end{center}
\caption{Sketches corresponding to the steps of the calculation of the elastic energy.
  {\bf (a)} Reference state (no deformation);  {\bf (b)} A point-load is applied at the
  free surface. The displacement at the application point is $\delta$. {\bf (c)} A sphere
  of radius $a$ indents the free surface over the distance $\delta$. {\bf (d)} Mapping
  from the reference state (solid lines, with the bead at the surface) to the deformed
  state. A point of the gel in the rest state is located with coordinates $(r,z)$.
  $\tilde{r}$ is the distance from the initial contact point of the bead. In the deformed
  state, the point that was at ($r,z$) is located at $R(r,z),Z(r,z)$.} 
\label{fig : scheme theory}
\end{figure}

The axis of symmetry being the vertical axis, the two coordinates changed by the deformation are
the radius in the horizontal plane, $r$, and the vertical coordinate, $z$. The deformation maps
 the undisturbed state with coordinates $(r,z)$ to a disturbed state $(R(r,z),Z(r,z))$, or,
in radial coordinates from coordinates $(\tilde{r};\theta)$ to $(R(\tilde{r},\theta);Z(\tilde{r},\theta))$
with $\tilde{r}$ radius and $\theta$ polar angle of the $(r;z)$ plane (see Fig.\ref{fig : scheme theory}-(d)).
The elastic energy is a function of the strain tensor (also called  {\em right Cauchy-Green deformation tensor}), which gives the square of local change in distances due to deformation: $C=F^TF$, $F$ being the deformation gradient tensor \cite{Ogden1984}. In the absence of any preferred direction, the elastic energy of an incompressible solid may depend only on two scalars (invariant under global rotation) that can be made out of the strain tensor: the trace ($I_1$) and the sum of the square of its components ($I_2$). In cylindrical coordinates with an azimuthal invariance, these invariants read $I_1=R_{,r}^2+\frac{R^2}{r^2}+R_{,z}^2+Z_{,r}^2+Z_{,z}^2$ and $I_2=(R_{,r}^2+Z_{,r}^2)^2+(R_{,z}^2+Z_{,z}^2)^2+2(R_{,r}R_{,z}+Z_{,z}Z_{,r})^2+\frac{R^4}{r^4}$, where indices preceded by a comma denote partial derivatives. For the neo-Hookean model, which is commonly used to describe soft gels, ${\cal W}=\frac{\mu}{2}I_1$. However, beyond a certain deformation, the neo-Hookean model cannot be a fair representation of rubber-like materials whose elasticity originates from unfolding of polymer chains: once these chains reach their full extension, the energy cost for a supplementary unfolding diverges so that a further deformation is accompanied by a diverging additional elastic energy, in contrast with the ideal neo-Hookean law (see Fig. \ref{fig : gent}). In what follows, we do not restrict ourselves to the neo-Hookean case.\\

 The deformation field of the gel can be described with two characteristic lengths, the downshift of the sphere $\delta$, and the radius of the sphere $a$. Here, we are dealing with the limit of large downshifts, that is  $\delta$ far larger than  $a$. Below we demonstrate that in this asymptotic case, the elastic energy of the system does not depend on $a$ in the  limit $a\ll \delta$.
 As a first step we consider a normal point-load applied on the free elastic surface at $(r,z)=(0,0)$ (Fig \ref{fig : scheme theory}-(b)) and we find a scaling law for the elastic energy per unit volume at a given distance $\tilde{r}$ from the loading point. The condition for the elastic energy to be convergent is established. Then as a second step  we consider a hard sphere with a finite radius, producing a displacement $\delta$ at the contact point $(r,z)=(0,0)$ equal to the displacement at $(r,z)=(0,0)$ induced by the previously considered point-like load. We demonstrate that for $\delta \ll a$, the elastic energy associated with the displacement field generated by the sphere or the point-like load are equal: These are independent of $a$ and proportional to $\delta^3$.

\subsection{Elastic energy for a point-like load} \label{sec : point}
Incompressibility of the elastic medium is imposed by writing that the determinant ${\cal D}$ of the first derivatives of ${\bf R}(x, y, z)$ is equal to one. For axisymmetric deformations, ${\cal D}=\frac{R}{r}\left(R_{,r}Z_{,z}-R_{,z}Z_{,r} \right)$. The total energy of the system \{gel + sphere\} reads :
\begin{equation}
 {\cal E}=\frac{4}{3}\pi \Delta \rho a^3 g Z(0,0)+2\pi\int_{-\infty}^0 \mbox{d}z \int_0^\infty \mbox{d}r \left( {\cal W}-q{\cal D}\right)r,
\label{eqn : full energy}
\end{equation}
where the Lagrange multiplier $q(r,z)$ imposes the incompressibility condition ${\cal D}=1$ \cite{Fox1987}.

 For the point-like heavy sphere we first consider (see Fig. \ref{fig : scheme theory}-(b)), the weight the bead can be reduced to a point force located at $(r,z)=(0,0)$. Eq. \ref{eqn : full energy} simplifies into:
\begin{equation}
 {\cal E}=\frac{16\pi^2}{3}\Delta \rho a^3 g\int_{-\infty}^0 \mbox{d}z \int_0^\infty r Z({\bf r})\delta^3({\bf r}) \mbox{d}r +2\pi\int_{-\infty}^0 \mbox{d}z \int_0^\infty \mbox{d}r \left( {\cal W}-q{\cal D}\right)r,
\label{eqn : energy}
\end{equation}
where $\delta^3 (r,z)$ is the 3D Dirac distribution. 

The Euler-Lagrange conditions of minimization of the energy (Eq. \ref{eqn : energy}) read \cite{Mora_prl2010}:
\begin{eqnarray}
  \left(\frac{\partial \left({\cal W}- q {\cal D}\right)r}{\partial R_{,r}}\right)_{,r}+ 
  \left(\frac{\partial \left({\cal W}- q {\cal D}\right)r}{\partial R_{,z}}\right)_{,z}&=&
  \frac{\partial \left({\cal W}- q {\cal D}\right)r}{\partial R}
  \label{eqn : cauchy r}\\ \nonumber  \\
  \left(\frac{\partial \left({\cal W}- q {\cal D}\right)r}{\partial Z_{,r}}\right)_{,r} + 
  \left(\frac{\partial \left({\cal W}- q {\cal D}\right)r}{\partial Z_{,z}}\right)_{,z}&=&\frac{8\pi r \Delta \rho g a^3}{3} \delta^3({\bf r}) \label{eqn : cauchy z}
\end{eqnarray}
Let $\delta$ be the displacement of the gel at $(r;z)=(0;0)$. $\delta$ is the unique relevant length for the displacement field. For points located at a distance to the load $\tilde{r}$ much smaller than $\delta$, $\delta$ is no more relevant at these short length-scales. Therefore no length-scales are expected to occur in the scaling laws and one assumes a power law of $\tilde{r}$ for the displacements for $\tilde{r}\ll \delta$:
$Z=\delta+\tilde{r}^{\beta} f_1(\theta)$ and $R=\tilde{r}^{\gamma} f_2(\theta)$. Since the gel
 is vertically stretched and horizontally squeezed in the vicinity of the bead one assumes that $ \gamma > \beta$. This hypothesis will be checked afterwards (see Eq. \ref{eqn : beta} below).

The constraint of conservation of volume reads (for $\tilde{r} \ll \delta$):
\begin{equation}
  {\cal D} =1 \Rightarrow  \overbrace{\frac{R}{r}}^{\sim \tilde{r}^{\gamma-1}} (\overbrace{R_{,r}}^{\sim \tilde{r}^{\gamma-1}}\overbrace{Z_{,z}}^{\sim \tilde{r}^{\beta-1}}-\overbrace{R_{,z}}^{\sim \tilde{r}^{\gamma-1}}\overbrace{Z_{,r}}^{\sim \tilde{r}^{\beta-1}})=1\Rightarrow \gamma = \frac{3-\beta}{2}. \label{eqn : gamma beta}
\end{equation}
The condition that $ \gamma > \beta$ with Eq. \ref{eqn : gamma beta}  yields $ \beta < 1$.\\

We assume that the strain energy density function is ${\cal W}\sim I_1^{\alpha_1} I_2^{\alpha_2} $ (the
exponents $\alpha_1$ and $\alpha_2$ are constant) in the range of strains undergone by the elastic solid
in the vicinity of the application point, {\em i.e.} for $\tilde{r}$ far smaller than $\delta$.  This choice for ${\cal W}$ is crucial to obtain the scaling law, but it has no effect on the final results, as argued in section \ref{sec : w}: it does not limit the general nature of the theory.\\

The scaling laws for the first and the second invariants of the Cauchy deformation tensor are:
\begin{eqnarray}
  I_1=Tr(C)&=&\overbrace{R_{,r}^2}^{\sim \tilde{r}^{1-\beta}}+\overbrace{\frac{R^2}{r^2}}^{\sim \tilde{r}^{1-\beta}}+\overbrace{R_{,z}^2}^{\sim \tilde{r}^{1-\beta}}+\overbrace{Z_{,r}^2}^{\sim \tilde{r}^{2\beta-2}}+\overbrace{Z_{,z}^2}^{\sim \tilde{r}^{2\beta-2}}\sim  \tilde{r}^{2\beta-2},
  \label{eqn : scaling I1}\\
  I_2=Tr(C^2)&=&(R_{,r}^2+Z_{,r}^2)^2+(R_{,z}^2+Z_{,z}^2)^2+2(R_{,r}R_{,z}+Z_{,z}Z_{,r})^2+\frac{R^4}{r^4}\sim \tilde{r}^{4\beta-4}.
  \label{eqn : scaling I2}
\end{eqnarray}
These scaling laws for $I_1$ and $I_2$ are used to obtain the scaling laws for the various terms of the Cauchy-Poisson equations (Eqs. \ref{eqn : cauchy r} and  \ref{eqn : cauchy z}). We first deal with Eq. \ref{eqn : cauchy r}:
\begin{eqnarray}
  \frac{\partial \left({\cal W}- q {\cal D}\right)r}{\partial R_{,r}}&=& r\frac{\partial {\cal W}}{\partial I_1}\frac{\partial I_1}{\partial R_{,r}}+ r\frac{\partial {\cal W}}{\partial I_2}\frac{\partial I_2}{\partial R_{,r}}-qr\frac{\partial {\cal D}}{\partial R_{,r} } \nonumber\\
  &=& \underbrace{2r\frac{\partial {\cal W}}{\partial I_1}R_{,r}}_{\sim \tilde{r}^{\frac{3-\beta}{2}+2(\alpha_1+2\alpha_2-1)(\beta-1)}}
  + \underbrace{4r\frac{\partial {\cal W}}{\partial I_2}\left(R_{,r}(R_{,r}^2+Z_{,r}^2)+R_{,z}(R_{,r}R_{,z}+Z_{,z}Z_{,r} \right)}_{ \sim \tilde{r}^{\frac{3-\beta}{2}+2(\alpha_1+2\alpha_2-1)(\beta-1)}}
  -q\underbrace{RZ_{,z}}_{\sim \tilde{r}^{\frac{\beta+1}{2}}}. \nonumber \\ \label{eqn : order 1}
\end{eqnarray}
In the same way one finds:
\begin{equation}
  \frac{\partial \left({\cal W}- q {\cal D}\right)r}{\partial R_{,z}}\sim \frac{3-\beta}{2}+2(\alpha_1+2\alpha_2-1)(\beta-1),
  \label{eqn : order 2}
\end{equation}

\begin{equation}
  \frac{\partial \left({\cal W}- q {\cal D}\right)r}{\partial R}\sim \tilde{r}^{\frac{\beta-1}{2}}. \label{eqn : order 3}
  \end{equation}
From Eq. \ref{eqn : cauchy r} and Eqs. \ref{eqn : order 1}, \ref{eqn : order 2} and \ref{eqn : order 3} one obtains $\tilde{r}^{\frac{1-\beta}{2}+2(\alpha_1+2\alpha_2-1)(\beta-1)} \sim  q \tilde{r}^{\frac{\beta-1}{2}}$,
and the scaling law for the Lagrange multiplier:
\begin{equation}
  q\sim \tilde{r}^{(\beta-1)(2\alpha_1+4\alpha_2-3)}.
  \label{eqn : scaling q}
\end{equation}

The last Cauchy-Poisson equation (Eq. \ref{eqn : cauchy z}) is now used to get an expression of $\beta$ as a function of $\alpha_1$ and $\alpha_2$. It writes:
\begin{equation}
\frac{1}{r}\left(r\frac{\partial \left({\cal W}- q {\cal D}\right)}{\partial Z_{,r}}\right)_{,r} + 
\left(\frac{\partial \left({\cal W}- q {\cal D}\right)}{\partial Z_{,z}}\right)_{,z}
=\frac{8\pi  \Delta \rho g a^3}{3} \delta^3({\bf r}).
\label{eqn : divergence}
\end{equation}
Denoting as $\sigma_{zr}=\frac{\partial \left({\cal W}- q {\cal D}\right)}{\partial Z_{,r}}$
and $\sigma_{zz}=\frac{\partial \left({\cal W}- q {\cal D}\right)}{\partial Z_{,z}}$, the two quantities present on the 
left-hand side of Eq.\ref{eqn : divergence}, one finds that this left-hand side is the divergence (expressed in cylindrical coordinates) of the z-components of the stress tensor $\sigma$ \cite{Ogden1984}. \\
We first calculate the scaling laws for $\sigma_{zr}$ and $\sigma_{zz}$ using Eqs. \ref{eqn : gamma beta}, \ref{eqn : scaling I1}, \ref{eqn : scaling q}:
\begin{eqnarray}
 \sigma_{zr}= \frac{\partial \left({\cal W}- q {\cal D}\right)}{\partial Z_{,r}}&=&  \frac{\partial {\cal W}}{\partial I_1}\frac{\partial I_1}{\partial Z_{,r}}+ \frac{\partial {\cal W}}{\partial I_2}\frac{\partial I_2}{\partial Z_{,r}}-q\frac{\partial {\cal D}}{\partial Z_{,r} }\\
  &=& \underbrace{2\frac{\partial {\cal W}}{\partial I_1}Z_{,r}}_{\sim \tilde{r}^{(\beta-1)(2\alpha_1+4\alpha_2-1)}}+q\underbrace{\frac{R}{r}R_{,z}}_{\sim \tilde{r}^{1-\beta}}.
  \label{eqn : compare q}
\end{eqnarray}
The second term of the right-hand side of Eq. \ref{eqn : compare q} scales as $\tilde{r}^{(\beta-1)(2\alpha_1+4\alpha_2-3)}\tilde{r}^{1-\beta}$. It is negligible when $ \tilde{r}$ tends to zero when compared to the first term of Eq. \ref{eqn : compare q} provided that the exponent of the second term is larger than the exponent of the first one, {\em i.e.}:
\begin{equation}
  (\beta-1)(2\alpha_1+4\alpha_2-3)+1-\beta>(\beta-1)(2\alpha_1+4\alpha_2-1),
\end{equation}
which is formally equivalent to $\beta<1$. Since it is assumed from the beginning that $\beta <1$, we conclude that $\sigma_{zr}\sim  \tilde{r}^{(\beta-1)(2\alpha_1+4\alpha_2-1)}$ and in the same way, $\sigma_{zz}\sim  \tilde{r}^{(\beta-1)(2\alpha_1+4\alpha_2-1)}$: $\sigma_{zr}$ and  $\sigma_{zz}$ follow the same power law when $ \tilde{r}$ tends to zero. The right-hand side of Eq. \ref{eqn : divergence} being a delta-like charge density, we conclude from Gauss's theorem that the scaling law for $\sigma_{zr}$ and  $\sigma_{zz}$ is also $\sigma_{zr}\sim \sigma_{zz}\sim \frac{1}{r^2}$. Therefore, $\tilde{r}^{(\beta-1)(2\alpha_1+4\alpha_2-1)}\sim \tilde{r}^{-2}$,
and:
\begin{equation}
  \beta=\frac{2\alpha_1+4\alpha_2-3}{2\alpha_1+4\alpha_2-1}.
  \label{eqn : beta}
\end{equation}
$\beta$ is an increasing function of $(\alpha_1+2\alpha_2)$. It is negative for $\alpha_1+2\alpha_2 <3/2$, and it is always smaller than $1$. The displacement at the application point is finite if the exponents are positive, {\em i.e.} if $\alpha_1+2\alpha_2>3/2$.
Otherwise the material cannot withstand a point-load. In the following we assume that $\alpha_1+2\alpha_2>3/2$, an ansatz that will be legitimized later.\\

Since $\delta$ is the unique length-scale of the deformation, the coordinates in the deformed configuration can be written as:
$R=\delta f_0\left(\frac{{\bf r}}{\delta}\right)$ and $Z=\delta g_0\left(\frac{{\bf r}}{\delta}\right)$, where
$f_0$ and $g_0$ are two numerical functions depending only on the constitutive law of the
elastic medium. The invariants $I_1$ and $I_2$, and thus the elastic energy density are dimensionless
functions depending only on ${\bf r}/\delta$. The elastic energy due to the deformation,
\begin{equation}
{\cal E}_{el~0}=\int\!\!\!\!\!\int\!\!\!\!\!\int_{z>0;\tilde{r}>0}\!\!\!\!\!\!\!\!\! {\cal W}_0 \mbox{d}^3{\bf r},
\label{eqn : energy 0}
\end{equation}
(subscripts 0 are for the point-load problem) is therefore proportional to $\delta^3$ with a
coefficient $C_0$, which has the dimension of shear modulus and depends on the mechanical
properties of the elastic solid, as:
\begin{equation}
  {\cal E}_{el~0} =  C_0 \delta^3.
  \label{eqn : energy 0 bis}
\end{equation}
Note that the convergence of the integral in Eq.\ref{eqn : energy 0} is ensured since the
work done by the applied force is finite for $\alpha_1+2\alpha_2>\frac{3}{2}$, yielding the validity condition of Eq. \ref{eqn : energy 0 bis}. 

\subsection{Elastic energy for a finite sphere}  \label{sec : finite}
\begin{figure}[!h]
  \begin{center}
    \includegraphics[width=0.45\textwidth]{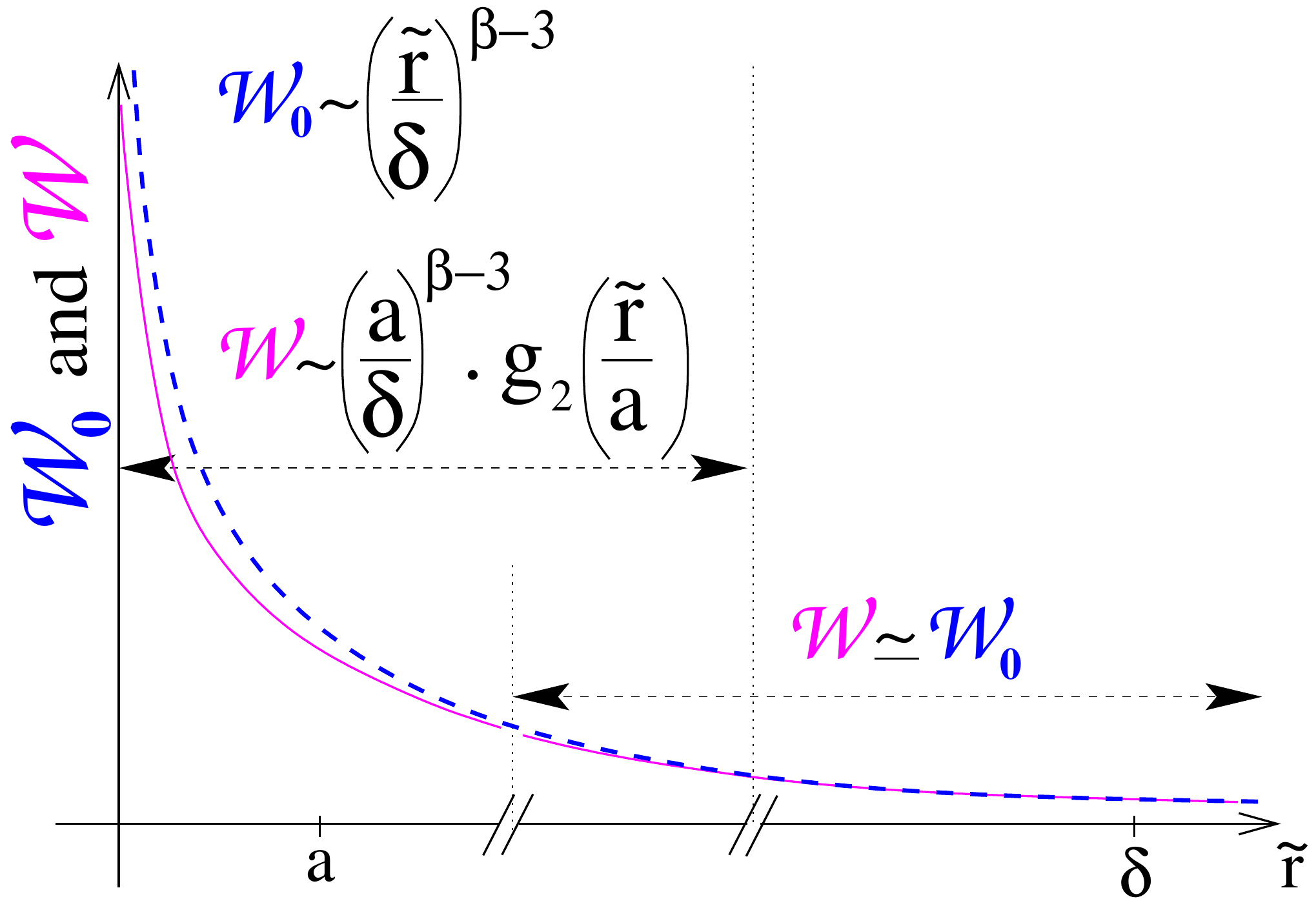}
  \end{center}
  \caption{Sketch of the energy density profiles ${\cal W}_0$ (e.g. dotted line) and ${\cal W}$
    (e.g. solid line) as a function of the distance t $\tilde{r}$ to the initial contact point. The dashed horizontal double arrows highlight two domains, corresponding to (i) $\tilde{r}\gg a$ where the elastic energy densities ${\cal W}$ and ${\cal W}_0$ are similar, and (ii) $\tilde{r} \ll \delta$ where they are different.} 
  \label{fig : energies}
\end{figure}

We consider now a heavy bead of radius $a$, and denote $\delta$ its vertical
downshift, which is supposed to be far larger than $a$ (Fig. \ref{fig : scheme theory}-(c)).
\begin{itemize}
\item{The unique relevant length-scale for $\tilde{r}\gg a$ being $\delta$, the displacement
  field reduces in this range of $\tilde{r}$ to the displacement field for a point-load with
  the same penetration depth $\delta$: $Z=\delta  g_0\left(\frac{{\bf r}}{\delta}\right)$, where $g_0$ has been defined in section \ref{sec : point}.}
\item{The power law for $Z-\delta$ with the exponent given by Eq. \ref{eqn : beta} applies in  the intermediate range $a\ll \tilde{r}\ll \delta$.  Combining it with the previous expression of $Z$ (valid for $\tilde{r}\gg a$) yields:
  \begin{equation}
    Z-\delta \sim \delta\left(\frac{\tilde{r}}{\delta}\right)^\beta = \delta \left(\frac{a}{\delta}\right)^\beta  \left(\frac{\tilde{r}}{a}\right)^\beta.
    \label{eqn : inter}
\end{equation}}
\item{Close to the bead ($\tilde{r}\ll \delta$) the unique relevant length-scale is $a$. $Z-\delta$
  can be expressed as 
\begin{equation} 
Z-\delta=Kg_1\left(\frac{{\bf r}}{a}\right),
\label{eqn : inter bis}
\end{equation} 
  where $g_1$ is a numerical
  function and the constant $K$ (with the dimension of a length) depends on $a$ and also on
  the far field deformation, {\em i.e.} on $\delta$. Comparing Eq. \ref{eqn : inter} with the general
  expressions for $Z$ in the range  $\tilde{r}\ll \delta$, one obtains
  $K=\delta \left(\frac{a}{\delta}\right)^\beta$.

 A similar expression for $R$ can be obtained with the exponent $(3-\beta)/2$, leading to negligible contributions for the elastic energy.}
\end{itemize}
For $\tilde{r}\ll \delta$, the first invariant $I_1$ scales as  $Z^2_{,z}$ and $Z^2_{,r}$, and the second invariant $I_2$ scales as $Z^4_{,z}$ and $Z^4_{,r}$.
 The strain energy density function ${\cal W}\sim I_1^{\alpha_1}I_2^{\alpha_2}$ can thus be written for  $\tilde{r}\ll \delta$ (using Eq. \ref{eqn : inter bis} with the expression of $K$) as  ${\cal W} \sim  \left(\frac{a}{\delta}\right)^{\beta-3}g_2(\frac{\bf{r}}{a})$. In the range $\tilde{r}\gg a$, ${\cal W}\simeq {\cal W}_0$, ${\cal W}_0$ being the elastic energy density of the point-load problem (Fig. \ref{fig : energies}).\\
Moreover, in the range $\tilde{r}\ll \delta$, one finds from the scaling laws of section \ref{sec : point}: ${\cal W}_0\sim \left(\frac{r}{\delta} \right)^{\beta-3}\sim \left(\frac{r}{a}\right)^{\beta-3}\left(\frac{a}{\delta}\right)^{\beta-3}$.\\ 

One concludes that, for any $\tilde{r}$, 
$
{\cal W}-{\cal W}_0 \sim \left(\frac{a}{\delta}\right)^{\beta-3}g_3(\frac{\bf{r}}{a}),
$
with $g_3$ a numerical function (depending neither on $\delta$ nor on $a$) whose limit as its
argument approaches $\infty$ equals 0. Thus, we obtain:
\begin{equation}
  \int\!\!\!\!\!\int\!\!\!\!\!\int_{z>0}\!\!\!\!\!\!\!\!\!\left({\cal W}-{\cal W}_0\right) \mbox{d}^3{\bf r}= \left(\frac{a}{\delta}\right)^{\beta-3}a^3 \int\!\!\!\!\!\int\!\!\!\!\!\int_{z>0}\!\!\!\!\!\!\!g_3\left(\frac{\bf{r}}{a}\right) \mbox{d}^3\left(\frac{{\bf r}}{a}\right) \sim  \delta^3\left(\frac{a}{\delta}\right)^{\beta}.
\label{eqn : scaling difference}
\end{equation}
Since one assumes again that $\alpha_1+2\alpha_2>\frac{3}{2}$, $\beta>0$ and the difference of the elastic energy with the bead of radius $a$,
${\cal E}_{el}$, to the elastic energy with the point-load, ${\cal E}_{el~0}$, is negligible
with respect to $\delta^3$. One concludes from Eq. \ref{eqn : energy 0 bis} that the elastic
energy  ${\cal E}_{el}$ is proportional to $\delta^3$.\\
Note that if $\alpha_1+2\alpha_2$ were smaller than $3/2$ the near-field part of the elastic energy would not be
negligible anymore compared to the far field contribution.\\

\subsection{The strain energy density function} \label{sec : w}

In sections \ref{sec : point} and \ref{sec : finite} the strain energy density function has been assumed to scale as ${\cal W}\sim I_1^{\alpha_1}I_2^{\alpha_2}$ for the strains encountered at $\tilde{r} \ll \delta$. It has been then demonstrated that the elastic energy corresponding to the downshift of a sphere of finite radius $a$ over the distance $\delta$ is proportional to $\delta^3$ in the limit $\delta \gg a$ if $\alpha_1+2\alpha_2>3/2$. \\

It is worth considering the case of an elastic material for which the strain energy density function increases softer than $I_1^{\alpha_1}I_2^{\alpha_2}$ with  $\alpha_1+2\alpha_2=3/2$.
For $\tilde{r}\ll \delta$ we demonstrated that  $Z-\delta=\delta \left(\frac{a}{\delta}\right)^\beta g_1\left(\frac{{\bf r}}{a}\right)$ and then $Z_{,z}\sim \left(\frac{\delta}{a}\right)^{1-\beta}$, showing that the strain is arbitrarily large in the limit $\delta \gg a$ , if $\beta<1$, {\em i.e.} if $\alpha_1+2\alpha_2<3/2$.
However, due to the finite maximum stretch of the polymer chains constituting the material, the maximum stretching of any real elastic rubber-like material is bounded. To this maximum stretching corresponds a divergence of the strain energy function. This divergence results in a steeper and steeper increase of ${\cal W}$ with $I_1$, $I_2$, or both, which is associated with increasing values of the exponents $\alpha_1$ and/or $\alpha_2$ \cite{Ogden1984}. 

\begin{figure}[!h]
  \begin{center}
    \includegraphics[width=0.45\textwidth]{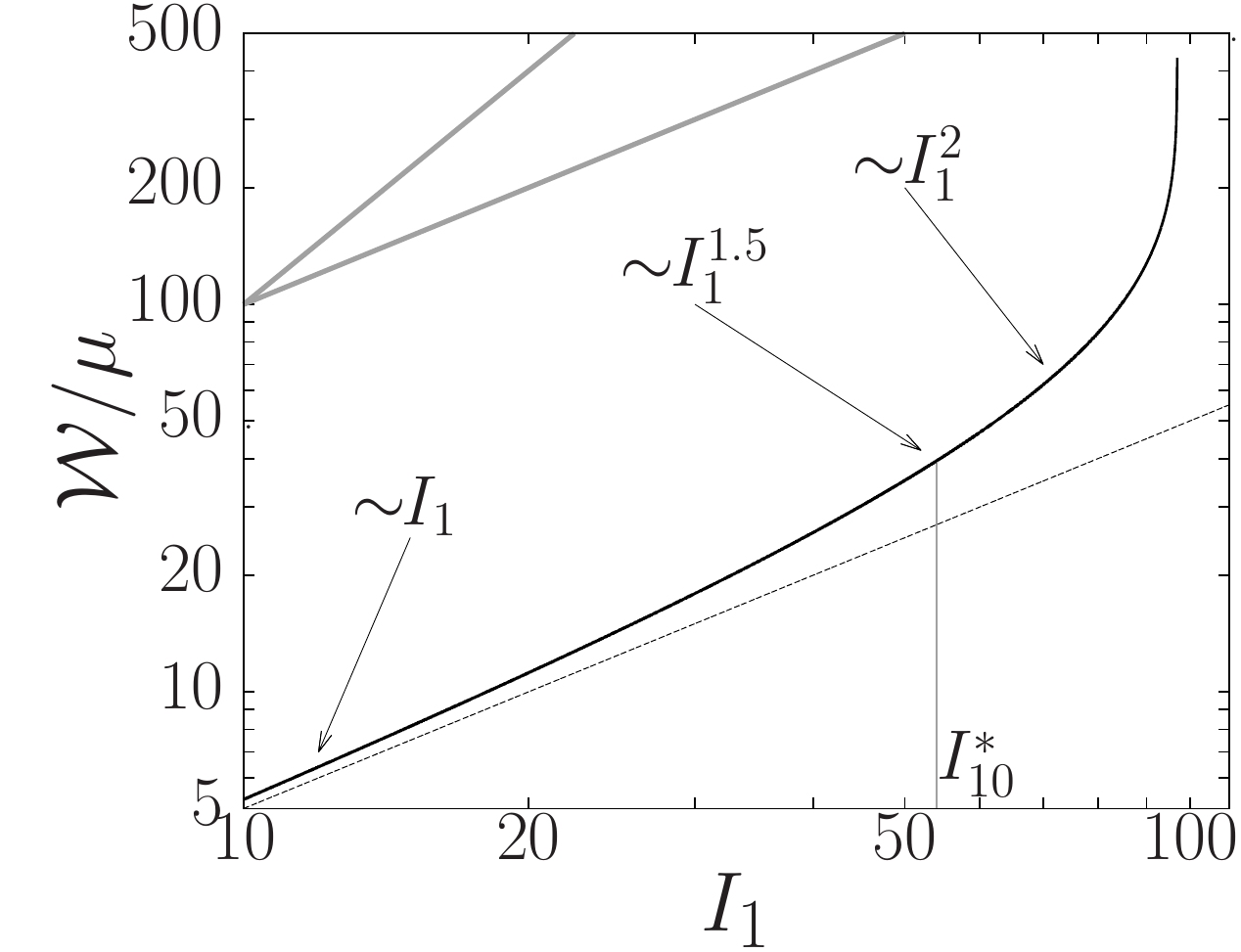}
  \end{center}
  \caption{Strain energy density function given by the incompressible Gent material model ${\cal W}=-\frac{1}{2}\mu J_m \log\left(1-\frac{I_1-3}{J_m} \right)$ \cite{Gent1996} with $J_m=97$, plotted in log-log scales (black solid line). $J_m+3$ is the  limiting value of the first invariant $I_1$.  The strain energy density function for the neo-Hookean material is plotted with dashed line. The two gray straight lines indicate slopes 1 and 2, {\em i.e.} values of the local exponent $\alpha_1$ equal to $1$ and $2$. $I_{10}^*$ is the value of the first invariant beyond which the local exponent is larger than $3/2$. } 
  \label{fig : gent}
\end{figure}

In order to illustrate this last point, let us consider a particular strain energy density function, for instance the Gent hyper-elastic model \cite{Gent1996,Gent2005}. The strain energy density function of this model has a singularity when the first invariant $I_1$ reaches a limiting value. It is plotted in Fig. \ref{fig : gent}: deviations from the initial neo-Hookean behaviour ($\alpha_1=1$ and $\alpha_2=0$) yield an increasingly stiffer and stiffer strain energy density function. At any value $I_{10}$ of $I_1$ one can define the {\em local} exponent $\alpha_{10}=\frac{I_1}{{\cal W}}\frac{\mathrm{d}{\cal W}}{\mathrm{d}I_1}$ so that at the vicinity of $I_{10}$ the strain energy density function is ${\cal W} \sim  I_1^{\alpha_{10}}$, locally. Beyond a certain finite value $I^*_{10}$ of $I_{10}$ the exponent is larger that $3/2$, ensuring that for any $I_1$ larger than $I^*_{10}$ the condition $\alpha_1+2\alpha_2>3/2$ is fulfilled. Note that values of $I_1$ larger than $I^*_{10}$ are automatically reached, otherwise the strain would diverge around $\tilde{r}\sim 0$, leading to arbitrarily high values of $I_1$ (as explained above). \\

What is illustrated using the example of the Gent model is general, and can be applied to any elastic constitutive law involving $I_1$ and $I_2$ of a real elastic material: starting from low to moderate strains for which the neo-Hookean model is expected to apply far from the bead, energy density functions stiffer than $I_1^{\alpha_1}I_2^{\alpha_2}$  with $\alpha_1+2\alpha_2> \frac{3}{2}$ are necessarily encountered next to the bead for any real elastic material.\\

 A material for which the failure limit is reached before the strain energy density is stiffer 
than $I_1^{\alpha_1}I_2^{\alpha_2}$ with $\alpha_1+2\alpha_2>3/2$ would not be able to sustain the heavy 
sphere, and thus would be drilled by the sphere.  These cases can be definitively excluded for the experiments we are dealing with since no fracture, plasticity nor creep have been observed during these experiments.

One concludes that for any elastic material, the elastic energy of the gel is proportional to $\delta^3$ within the limit $\delta \gg a$. The proportionality constant is $C_0$ (see Eq. \ref{eqn : energy 0 bis}) depends on the whole elastic constitutive law of the elastic material.   

\subsection{Equilibrium condition}

The energy is minimum in the equilibrium state. The $\delta$-derivatives of ${\cal E}_{gr}$ and ${\cal E}_{el}$ are therefore equal giving the scaling law valid for $\delta\gg a$, for a given elastic material:  
\begin{equation} 
  \delta \sim a^{3/2}\left(\Delta \rho g\right)^{1/2}.
\end{equation}
Note that the beads has been assumed to be spherical although this is not a crucial point in this theory: if it is compact with all typical lengths of the same order $a$ the previous scaling arguments apply.

\section{Discussion and conclusion}

The theoretical analysis presented above provides an understanding {\em sui generis} of the
experimentally observed scaling law.\\

The elastic material being chosen (so that the proportionality constant $C_0$ is fixed), this law predicts the scaling behaviour of the downshift of spheres with different radii and/or different densities. These predictions are in quantitative agreement with the observations: one observes experimentally a scaling law with an exponent close to $3/2$ for the radius, and the prefactor is not only related to the linear elastic properties ($\mu$), but also to non-linear features of the elastic material. These non-linear features being {\em a priori} distinct from one gel to another one, this explains why plotting $\delta/\delta_0$ as function of $a/\delta_0$ for different gel compositions yields to different curves.  The prefactor in the scaling law depends not only on
the shear modulus of the solid, but also on the elastic behaviour at large deformations,
in agreement with the experiments (\cite{Crosby2015}).
This provides a way to assess some characteristics of elastic 
materials under large strains, as strain stiffening properties.\\ 

The range of applications of the derived scaling goes far beyond the description of the elasto-buoyancy phenomenon. For instance, it can be directly checked that the recent indentation experiments performed on various types of compliant gels, by Fakhouri et al \cite{Crosby2015}, are in quantitative agreement with our prediction, however their very naive model based on the neo-Hookean equation, misses the driving non-linearities.\\

In this paper, we have shown that the exponent $3/2$ for the radius is independent of
the strain-stress relation provided that the increase in the elastic energy density
(${\cal W}$) with the strain is stiff enough. Generic behaviors for elastic materials 
undergoing large and complex deformations can therefore be identified, going beyond 
scaling arguments that is blind to the crucial effect of strain stiffening. This work sheds new light on the mechanics of extremely large elastic deformations that are crucial in various emerging techniques that even encompass an important procedure such as the computer-assisted surgery involving human organs \cite{Cotin1999,Liu2003,Taylor2008}, which are indeed non-linear materials and undergo large deformations. In all these important fields, an ideal neo-Hookean model is used that, according to our current work, is not valid {\em sensu stricto}. The danger is that one may miss the divergences of the solution that is inherent in the ideal neo-Hookean model because the full solution is not used. The work presented here is a panacea to the possible pitfalls that one may embark upon in studying extreme deformations of non-linear materials, thereby also offering possible benchmarks for numerical simulations and even opportunities to formulate new simulations methods.


\end{document}